\newcommand{\NVO}{NaV$_2$O$_5$}
\newcommand{\CGO}{CuGeO$_3$}
\newcommand{\VFOUR}{V$^{4+}$}
\newcommand{\VFIVE}{V$^{5+}$}
\documentstyle[prl,aps,epsf,twocolumn]{revtex}
\large
\begin{document}
\draft
\twocolumn[\hsize\textwidth\columnwidth\hsize\csname
@twocolumnfalse\endcsname

\title{Charge ordering and opening of spin gap in NaV$_2$O$_5$}
\author{
M. V. Mostovoy\cite{Perm1} and D. I. Khomskii\cite{Perm2}}
\address{Theoretical Physics and Materials Science Center,\\
University of Groningen, Nijenborgh 4, 9747 AG
Groningen, The Netherlands}
\date{\today}
\maketitle
\begin{abstract}
\widetext
\leftskip 54.8pt
\rightskip 54.8pt

We suggest that the phase transition observed in \NVO\ at $T = 34K$ is
not a spin-Peierls transition, but a charge ordering
transition, related to the formal presence in this system of
equal number of \VFOUR\ and \VFIVE\ ions.  Below $T_c$, \VFOUR\
ions form a zigzag structure, which is consistent with the
experimentally observed doubling of the lattice period in $a$
and $b$ directions.  We show that this charge ordering also
results in the alternation of spin exchange constants along the
$b$-direction, which opens a gap in the spin excitation
spectrum.  We emphasize the role of lattice distortions
around V ions both in the formation of the charged ordered state
and in the spin-gap opening.

\leftskip 54.8pt
\end{abstract}

]

\narrowtext

In this paper we discuss the interplay between charge and spin
degrees of freedom in NaV$_2$O$_5$, which initially was
identified as an inorganic spin-Peierls (SP) material similar to
\CGO.  Indeed, according to the first X-ray studies of this compound
\cite{Galy75} \VFOUR\ ions form spin-$\frac12$ chains, which
are separated from each other by nonmagnetic
\VFIVE-chains.  Below $T_c = 34$K, magnetic susceptibility, X-ray
and neutron scaterring measurements indicate the opening of a
spin gap accompanied by doubling of the lattice period in the
chain direction \cite{Isobe96,Fujii97,Weiden97}.

However, the situation in \NVO\ is evidently more complicated
and more interesting than in conventional SP systems.
First of all, the original crystal structure was recently
questioned \cite{Weber,Schnering,Palstra}.  According to the new
X-ray and neutron data, the structure of \NVO\ was
identified as a centrocymmetric group $Pmmn$, in contrast to the
earlier assignement $P2_1mn$ \cite{Galy75}.  The new structure
implies that all V sites are equivalent.  As an average valence
of V in \NVO\ is $4.5+$, there is one $d$-electron per two
equivalent V ions, which makes the description of this compound
as a spin-$\frac12$ chain material not obvious.

Other indications that the physics of \NVO\ may be different from
that of, e.g., \CGO\ come from thermodynamic data.  In
particular, the study of specific heat \cite{Buechner} shows that
the entropy of the transition in \NVO\ is larger than the entropy
of a pure SP system.  The ratio $\frac{2 \Delta_0}{T_c}$, where
$\Delta_0$ is the value of the spin gap at $T=0$, is for \NVO\
$\sim 6$ -- much larger than for the known SP materials, where it
is close to $3.5$,  which follows from the weak-coupling mean field
theory of SP transition \cite{Bulaevskii78}.  Another evidence
that the phase transition in \NVO\ may be of different nature
comes from a rather weak dependence of $T_c$ on magnetic field:
the shift of $T_c$ is $\sim 5$ times smaller
\cite{Buechner,Kremer} than the theoretical predictions
\cite{Bulaevskii78,Cross79}.  Yet another spectacular difference
in the behavior of \NVO\ and \CGO\ is the temperature-dependence
of the thermal conductivity \cite{Vasil'ev}: below $T_c$ the
thermal conductivity of \NVO\ increases by a factor of $5$, while
in \CGO\ the corresponding anomaly is very weak.

\begin{figure}
\centering \leavevmode
\epsfxsize=7cm
\epsfbox{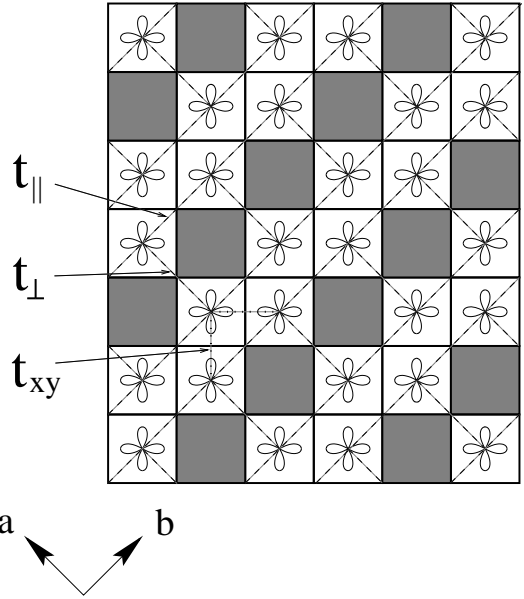}
\caption{
\footnotesize
Schematic representation of the crystal structure of V-O plane of
\NVO. Oxygenes are located at the corners of plaquettes and
vanadium ions are located at their centers; the shaded plaquettes
are vacant. Also shown are the relevant $d_{xy}$-orbitals of V
ions and vanadum ladders (dashed lines).
\label{VOlayer}}
\end{figure}

All these facts show that the phase transition in \NVO\ is not an
ordinary SP transition.  We suggest below that the main
phenomenon responsible for the transition is, in fact, a charge
ordering (CO), related to the formal presence in this system of
equal number of \VFOUR\ and \VFIVE\ ions.  While at high
temperature there is a rapid oscillation between the \VFOUR\ and
\VFIVE\ states, so that on the time scale larger than the
electron hopping time all $V$ ions look equivalent (V$^{4.5+}$),
at low temperature a CO transition, similar to the Verwey
transition in Fe$_3$O$_4$ \cite{Mott}, occurs. The opening of a
spin gap, acompanying this transition, is just one of the
consequences of the charge ordering.  The appearance below $T_c$
of two ineqivalent V sites, which were crudely identified as
\VFOUR\ and \VFIVE, was indeed observed in the recent NMR
experiment \cite{Ohama}.  However, the detailed nature of this CO
and its relation to the opening of the spin gap remains still
unclear.  Elucidation of this relation is the main subject of
this paper.

The structure of \NVO\ consists of V$_2$O$_5$ layers with
approximately square oxygen lattice.  Two third of the oxygen
plaquettes are occupied by V ions located in the middle of the
plaquettes (see Fig.~1). Due to details of the crystal structure,
only one of the three $t_{2g}$-orbitals of V, namely the
$d_{xy}$-orbital, is occupied \cite{Weber}.  As the $d$-$d$
hopping goes mainly via intermediate oxygens, the diagonal (or
next-nearest-neighbor (nnn)) hopping amplitudes, $t_{\perp}$ and
$t_{\parallel}$, are much larger than the nearest-neighbor (nn)
hopping amplitude $t_{xy}$, which makes \NVO\ a quater-filled
system of vanadium two-leg ladders (See Fig.~\ref{VOlayer}).
Tight binding fitting of the LDA band structure gives $t_{\perp}
\sim 0.35$eV, $t_{\parallel} \sim 0.15$eV, and $t_{xy} \sim
0.012$eV \cite{Weber}, all remaining hopping amplitudes being
much smaller.  Using the empirical Harrison rules Horsh and Mack
obtained similar values for $t_{\perp}$ and $t_{\parallel}$, but
a much larger $t_{xy} \sim 0.3$eV \cite{Horsh}.

With the parameters of Ref.~\cite{Weber} one can understand the
one-dimensional spin structure of \NVO.  Each rung is occupied by
a single electron in a bonding state with the energy
$-t_{\perp}$.  For strong on-site Coulomb repulsion the energy
cost of electron transfer on a neighboring rung is (at least) $2
t_{\perp}$.  For $2 t_{\perp} > 4 t_{\parallel}$ the electron
hopping between rungs is energetically unfavorable and can occur
only virtually.  Thus, although the charge degrees of freedom are
not frozen and all V ions are equivalent, the system remains
insulating.  In Refs.~\cite{Weber,Horsh} \NVO\ was described as
an effective spin system with spins being localized on the rungs
of vanadium ladders (instead of vanadium sites).  As the
spin-exchange between electrons of neighboring ladders is weak
\cite{Weber,Horsh}, the spin system is quasi-one-dimensional
(each ladder corresponding to a spin chain).

In this paper we go beyond the decription of \NVO\ as a
spin-chain material and include also the charge dynamics.
We start from the electronic Hamiltonian, which includes the
electron hopping in ladders (with the amplitudes $t_{\perp}$ and
$t_{\parallel}$) and between ladders (with the amplitude $t_{xy}$)
[see Fig.~\ref{VOlayer}], as well as the Coulomb interaction
between electrons on different V sites: the on-site Coulomb
repulsion $U$ is the largest parameter and is here taken
infinite, the nn interaction $V_2$ (e.g., between
sites C and D on neighboring ladders [see Fig.~\ref{alter}]), the
nnn interaction $V_1$ inside ladders (e.g., between sites A and E
or B and E [see Fig.~\ref{alter}]), etc.

Next we make a projection on the subspace of states with one
electron per rung.  Four different states of a single electron on
a rung can be represented as the eigenstates of spin $S =
\frac12$ and isospin $T = \frac12$ operators, $| \frac12 S^z
\rangle \otimes | \frac12 T^z \rangle$, where $T^z = \pm\frac12$
corresponds to an upper/lower position of the electron on a rung
(see Fig.~\ref{alter}).

The effective spin-isospin Hamiltonian can then be written in the
form:
\begin{equation}
H = H_0 + H_1.
\label{ham}
\end{equation}
Here, $H_0$ is a pure isospin Hamiltonian:
\begin{eqnarray}
H_0  =  &-& 2 t_{\perp} \sum_{\bf r} T^x_{\bf r} +
2 V_{1} \sum_{\bf r}  \left(T^z_{\bf r} T^z_{{\bf r}+ {\bf b}} +
\frac14\right) \nonumber \\
&+& 2 V_2 \sum_{\langle {\bf r}_1 {\bf r}_2\rangle}
\left(\frac14 - T^z_{{\bf r}_1} T^z_{{\bf r}_2}\right) +
H_0^{\prime},
\label{H_0}
\end{eqnarray}
where the vector ${\bf r}$ runs over the sites of an effective
lattice located on the rungs of V ladders.  The first term in
Eq.(\ref{H_0}) describes the electron hopping in the rungs,
the second term is the nnn Coulomb interaction between electrons
occupying neighboring rungs ${\bf r}$ and ${\bf r} + {\bf b}$ of
a ladder (${\bf b}$ is a unit lattice vector in the ladder
direction), the third term is the repulsion between nn electrons
from neighboring ladders and $H_0^{\prime}$ stands for all other
interaction terms (the longer range Coulomb terms, the
interaction via the lattice, etc.).

The Hamiltonian $H_1$ in Eq.(\ref{ham}) decsribes the
interaction between spin and isospin degrees of freedom. The
strongest spin-isospin interaction occurs inside ladders. The
corresponding Hamiltonian for a single ladder has the form:
\begin{eqnarray}
H_{ST}  = \!&-&\! \frac{4 t_{\parallel}^2}{\Delta}\!
\sum_n\! \biggl[\!
\biggl(\!\frac14 \!-\! {\bf S}_n \!\!\cdot\! {\bf S}_{n + 1}\!\!\biggr)\!\!
\biggl(\!\frac14 \!+\! {\bf T}_n \!\!\cdot \!{\bf T}_{n + 1}
\!-\! 2 T_n^z T_{n+1}^z \!\!\biggr) \nonumber \\
\!&+&\! \biggl(\frac34 \!+\! {\bf S}_n \!\!\cdot\! {\bf S}_{n + 1} \biggr)\!
\biggl(\frac14 \!-\! {\bf T}_n \!\!\cdot\! {\bf T}_{n + 1} \biggr)
\biggr],
\label{H_1}
\end{eqnarray}
where $n$ runs over rungs of the ladder.

The Hamiltonian Eq.(\ref{H_1}) originates from the virtual
electron hopping between two neighboring rungs of a ladder and
$\Delta$ is the increase of energy due to electron transfer on
the neighboring rung.  The value of $\Delta$ is determined
by the isospin Hamiltonian $H_0$ and depends on a state considered.
If, for instance, we neglect the Coulomb terms in the disordered
phase, then $\Delta = 2 t_{\perp}$, while for the extreme zigzag
CO discussed below: $\Delta \sim 2 V_1$. Note, that in the $U
\rightarrow \infty$ limit $\Delta$ is finite. Since the hopping
between ladders, $t_{xy}$, is much smaller than $t_{\perp}$ and
$t_{\parallel}$ \cite{Weber}, we ingnored the corresponding
contributions to $H_1$.

The spin-isospin structure of the exchange interaction is a
consequence of the Pauli principle: the wave function of two
neighboring electrons in the intermediate state (when they occupy
the same rung) should be antisymmetric under exchange of both
their spin and isospin coordinates.  Therefore, if the total spin
of two electrons on neighboring rungs, ${\bf S} = {\bf S}_n +
{\bf S}_{n+1}$, is zero (i.e., the spin state is antisymmetric),
the isospin state of these electrons has to be symmetric: $T =
1$, where ${\bf T} = {\bf T}_n + {\bf T}_{n+1}$, and, vice versa,
if $S = 1$, then $T = 0$.  Correspondingly, the first term in
square brackets of Eq.~\ref{H_1} is the projection operator on $S
= 0$, $T = 1, T^z = 0$ state, while the second term is the
projection operator on $S = 1$, $T = 0$ state.  The projection
of the total isospin of two electrons, $T^z$, in both terms
is zero, because we assumed the on-site Coulomb repulsion
$U$ to be infintely large, so that the hopping between rungs is
only possible if electrons are located on different chains of the
ladder.

\begin{figure}
\centering \leavevmode
\epsfxsize=7cm
\epsfbox{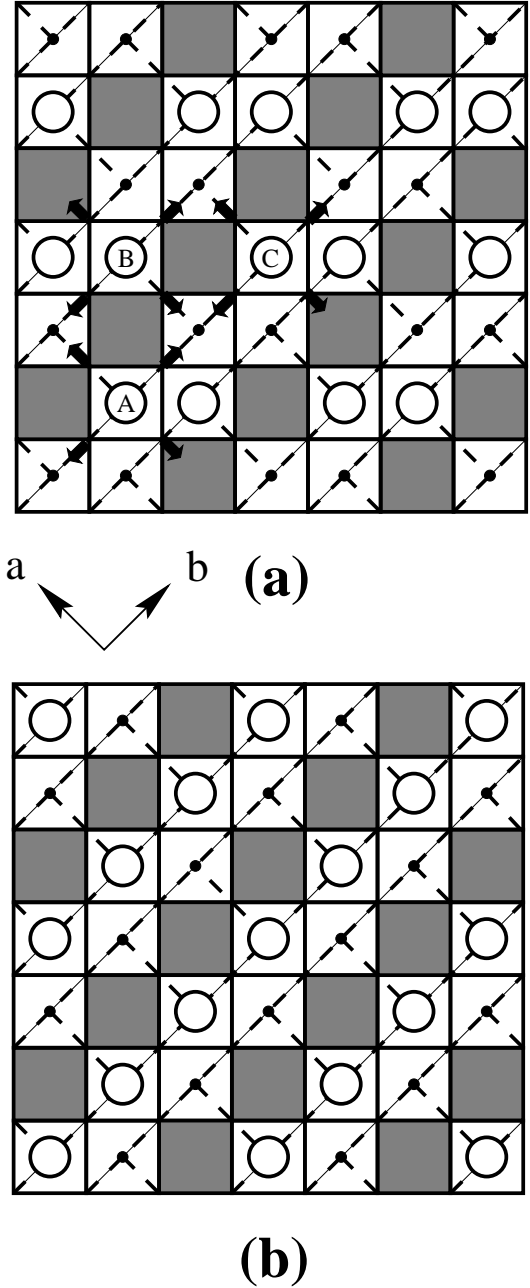}
\vspace{2mm}
\caption{
\footnotesize
Two types of CO: the zigzag structure (a) and the old structure,
in which \VFOUR\ chains are separated by \VFIVE\ chains (b).
Large circles denote \VFOUR\ ions with a large ionic radius,
while small circles denote small nonmagnetic \VFIVE\ ions.
Arrows denote shifts of oxygen atoms around \VFOUR\ ions in one
ladder (for simplicity, the shifts induced by \VFOUR\ ions in
other ladders are not shown).
\label{chord}}
\end{figure}

Since $H_1$, resulting from the virtual hopping between rungs, is
smaller than $H_0$, we can start by considering only the charge
degrees of freedom first.  The isospin Hamiltonian $H_0$
describes the competition between the hopping along rungs, which
tends to make V ions equivalent, and the Coulomb interaction,
which favors a CO.  If, for instance, we leave only the first two
terms in Eq.(\ref{H_0}), then the ladders become decoupled and
the Hamiltonian of each ladder is the Hamiltonian of the Ising
model in perpendicular magnetic field.  For $V_1 < 2
t_{\perp}$ its ground state is disordered (all vanadium ions are
equivalent), while in the opposite case, isospins in the ground
state are ordered antiferromagnetically: $\langle 2 T^z_n \rangle
= \eta (-1)^n$.  This CO corresponds to the zigzag occupation of
vanadium sites by electrons as shown in Fig.~\ref{chord}a , where
the large circles denote \VFOUR\ ions with a large ionic radius,
while small circles denote small nonmagnetic \VFIVE\ ions.  (Of
course, if the ladders would be really decoupled, the sign of the
order parameter in each ladder is arbitrary.)

On the other hand, the third term in Eq.(\ref{H_0}) favors the
structure shown in Fig.~\ref{chord}b , which would give
one-dimensional spin chains of $V^{4+}$ ($d^1$, $S = \frac12$)
ions and is the initially assumed crystal structure
\cite{Galy75}.  If we would take into account only the second and
the third terms with $V_2 = \sqrt{2} V_1$, the zigzag structure
(Fig.~\ref{chord}a) would have lower energy than the chain
structure (Fig.~\ref{chord}b).  Of course, it is necessary to
include all longer range Coulomb interactions to find the real
ground state of the system.  The calculation of the Madelung
energies of these two-dimensional structures shows, that they are
rather close, the chain structure of Fig.~\ref{chord}b being
slightly more favorable.

There is, however, another type of interaction
between charges -- the interaction via lattice distortions,
which, unlike the Coulomb interaction, is local.  As shown in
Fig.~\ref{chord}a, this latter interaction definitely favors the
zigzag structure: large $V^{4+}$ ions push out neighboring
oxygens (the directions of oxygen displacements in a ladder are
shown in Fig.~\ref{chord}a by arrows).  As a result, neighboring
plaquettes in a ladder would preferably be occupied by smaller
$V^{5+}$ ions.  Roughly speaking, the lattice relaxation around
empty and occupied vanadium ions enhances the nnn repulsion,
which favors the zigzag ordering.

Note, that the zigzag structure shown in Fig.~\ref{chord}a
immediately gives us a new periodicity with the doubling of
periods along both $a$ and $b$ directions, in accordance with the
experiment \cite{Fujii97}.  Thus, one does not need to invoke the
SP mechanism to explain the distortion pattern observed below
$T_c$. We also note, that there exist four equivalent
realizations of the zigzag structure of Fig.~\ref{chord}a, which
differ by the location and orientation of pairs of nn \VFOUR\
ions (large circles in  Fig.~\ref{chord}a). This may provide a
clue to understanding of the four-fold increase of the period in
$c$-direction below $T_c$ \cite{Fujii97}.

The smallness of $T_c = 34K$ in \NVO\ as compared to the
parameters of the Hamiltonian (\ref{H_0}) (e.g., $t_{\perp}$,
$V_1$) may be explained by the competition between the electron
hopping and the electron-electron interactions, which reduces the
gap in the spectrum of isospin excitations. If, again, we
would leave only the first two terms in the Hamiltonian
(\ref{H_0}), the gap equals $|2 t_{\perp} - V_1|$. Although these
excitations do not carry charge, they can be excited by an
electric field applied perpendicularly to ladders (i.e., in
$a$-direction), which may explain why the continuum of relatively
low-energy excitations, observed in the infrared absorption
experiments, disappears when the orientation of the electric
field is changed from $a$-direction to $b$-direction.
\cite{Dirk}. The repulsion between electrons from neighboring
ladders (the third term in Eq.(\ref{H_0})) frustrates the zigzag
structure, which results in additional reduction of $T_c$.

Next we consider the spin excitations. From Eq.(\ref{H_1})
the effective spin-exchange constant in the chain direction,
$J_{\parallel}$, is
\begin{equation}
J_{\parallel} = \frac{4 t_{\parallel}^2}{\Delta}
\langle {\bf T}_n \!\!\cdot\! {\bf T}_{n + 1} \!-\! T^z_n T^z_{n
+ 1} \rangle = \frac{4 t_{\parallel}^2}{\Delta}
\langle T^x_n T^x_{n+1} \!+\!
T^y_{n} T^y_{n+1} \rangle .
\label{J}
\end{equation}
where $\langle \ldots \rangle$ denotes the thermal and quantum
average.

From Eq.(\ref{J}) it is obvious that CO would decrease the
effective spin-exchange constant $J_{\parallel}$.  In particular,
if the electrons in neighboring rungs $n$ and $n+1$ of a ladder
are completely uncorrelated, $\langle T^x_n T^x_{n + 1} + T^y_n
T^y_{n + 1} \rangle = \frac12$, while for the extreme zigzag
charge ordering ( $\langle 2 T^z_n \rangle = (-1)^n $): $\langle
{\bf T}_n \cdot {\bf T}_{n + 1} - T^z_n T^z_{n + 1}
\rangle = 0$ and the spin-exchange constant is zero.  In general,
the charge order parameter $\eta$ does not have the maximal
amplitude, so that CO would result only in a partial reduction of
$J_{\parallel}$.  According to Eq.(\ref{J}), the spin stiffness
in the chain direction becomes smaller as temperature decreases,
which could be observed in neutron scattering experiments.

\begin{figure}[htbp]
\centering \leavevmode
\epsfxsize=8cm
\epsfbox{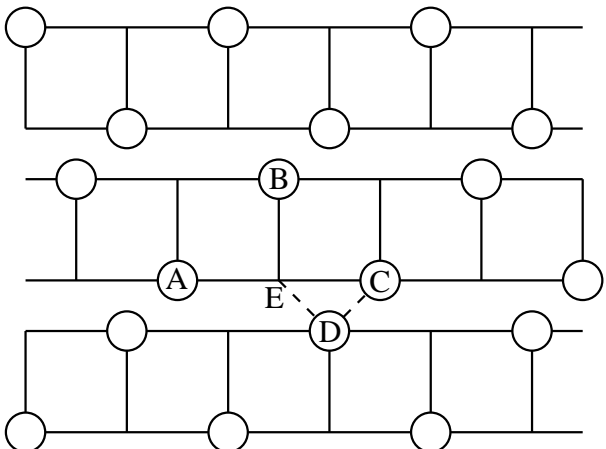}
\caption{
\footnotesize
This figure demonstrates that the zigzag CO makes the
spin-exchange in neighboring pairs of electrons ($AB$ and $BC$)
inequivalent. The inequivalence is caused by the difference in
the occupation of V sites in neighboring ladders.
\label{alter}}
\end{figure}

More importantly, the zigzag charge ordering opens a spin gap,
which in this picture is a secondary effect.  There are, in
principle, two scenarios for the spin gap opening.  First, as one
can see from Fig.~\ref{chord}a, the CO results in the appearance
of pairs of nn vanadiums coupled by the hopping amplitude
$t_{xy}$, [cf.  Fig.~\ref{VOlayer}].  For large $t_{xy}$ these
interladder pairs would form spin singlets, which could explain
the spin gap in \NVO\ \cite{brick}.  However, we think that this is
rather unlikely.  On the one hand, as we mentioned above, the
band structure calculations give rather small value of $t_{xy}
\sim 0.012$eV \cite{Weber}.  On the other hand, even with larger
$t_{xy}$ (as, e.g., in Ref.~\cite{Horsh}) there are several
competing mechanisms of nn exchange: an antiferromagnetic one,
due to direct d-d overlap and an exchange via $90^{\circ}$ V-O-V
path, which according to Goodenough-Kanamori-Anderson rule is
ferromagnetic.  Even the sign of the resulting interaction is
unclear: Horsh and Mack obtained a small antiferromagnetic nn
exchange \cite{Horsh}, while the LDA+U calculations for the
structurally similar material CaV$_2$O$_5$ give a ferromagnetic
nn interaction \cite{Korotin}.  Furthermore, in \NVO\ there is a
third mechanism of exchange resulting from the circular motion of
electrons over V-V-V triangles (e.g., along the CDE triangle in
Fig.~\ref{alter}).  The sign of the latter exchange coincides
with the sign of the product of three amplitudes of electron
hopping along the sides of a triangle [cf. \cite{Eskes}].  As the
later sign is negative \cite{Weber}, the corresponding
spin-exchange is ferromagnetic.

Thus, whether the interladder singlets are formed or not, is
still not clear.  It is, however, obvious that the zigzag CO
results in an alternation of the exchange constants {\em along}
the spin chains formed by vanadium ladders.  Indeed, the exchange
interaction between the ions A and B (see Fig.~\ref{alter}) goes
via a plaquette having $V^{5+}$ both above and below it, whereas
for $B$ and $C$ ions the corresponding positions are occupied by
$V^{4+}$ ions.  This inequivalence in the occupation of V sites
in neighboring ladders produces a difference in the shifts of the
corresponding oxygen and vanadium ions, thereby affecting the
electron hopping amplitudes and, ultimately, resulting in the
alternation of the exchange constants along the ladder direction,
\begin{equation}
J_{\parallel}(n,n+1) =
J_{\parallel} (1 + (-1)^n \delta),
\end{equation}
which opens the spin gap $\sim \frac{\delta^{2/3}}{\sqrt{|\ln
\delta|}}$. The detailed calculation of the alternation amplitude
$\delta$ induced by the CO is a formidable problem requiring the
knowledge of the shifts of oxygen and vanadium ions, the
modification of electron transfer energies, etc. One can,
however, easily see, that $\delta$ is a linear function of the
charge order parameter $\eta = (-)^n\langle 2 T^z_n \rangle$, so
the temperture behavior of the spin gap is quite similar to that
in SP systems, where $\delta$ is proportional to the amplitude of
the lattice distortion (unless, of course, the CO transition is
of the first order).

Though the opening of the spin gap is not the main driving force
of the transition, it also contributes to the stabilization of
the charge ordered phase.  It may, in particular, be important in
determining the relative phases of charge order parameter $\eta$
in different ladders. If, for instance, for the phase of the
charge order in the bottom ladder in Fig.~\ref{alter} would be
opposite, there would no doubling of the periodicity in
$a$-direction and no alternation in the spin-exchange constants
along the ladders, and consequently, no spin gap.  The energy
gain due to spin-gap opening, by itself, can make the structure
of Fig.~\ref{chord}a more stable than the zigzag structure
without the dimerization along $a$-axis.

Summarizing, we have shown that the very chemical composition of
\NVO\ -- the presence of one electron per two vanadium sites --
rather naturally leads to a charge ordering at low temperatures.
We argued, that in the ordered phase the occupied vanadum sites
form zigzags.  This CO is driven by the direct Coulomb
repulsion and, most importantly, by the interaction with the
lattice distortions.  The CO results in the doubling of the
lattice period both in $a$ and $b$ directions in agreement with
experiments.  Another consequence of the CO is the alternation of
the exchange interaction along $b$-direction (along vanadium
ladders), which openins a spin gap.  Although the coupling
between spin and charge degrees of freedom may also contriubte to
the charge ordering, the spin-gap opening is a consequence of the
charge ordering.  This picture is consistent with experimental
observations for \NVO.

We are very grateful to A.~Damascelli, R.~K.~Kremer, D.~van der
Marel, T.~M.~Rice, W.~Schnelle, A.~Smirnov, A.~N.~Vasil'ev, and
especially to B.~B\"uchner for useful discussions.  This work was
supported by the ``Stichting voor Fundamenteel Onderzoek der
Materie (FOM)'' and by the European Community through the network
OXSEN.

\vspace{5ex}

\noindent
{\em Note added: }
After completion of this work two preprints appeared
\cite{Fukuyama,Fulde} in which similar ideas were put forward.
Thalmeier and Fulde \cite{Fulde} argued in favor of the CO shown
in Fig.~\ref{chord}b, after which they still had to invoke the
spin-lattice coupling to get second transition of a
spin-Peierls type. Experimentally two close phase transitions
were observed in some samples of \NVO\ \cite{Koeppen} but not in
others \cite{Kremer}. Furthermore, when two transitions are
observed, their temperature shifts, essentially, in the same way
with magnetic field \cite{Buechner}.  Thus, it may well be that
these multiple transitions originate from the inhomogeneity of some
samples, although this question definitely deserves further
study. However, as we argued above, the zigzag structure of
Fig.~\ref{chord}a seems to us much more favorable.  Besides, in
the model of Ref.~\cite{Fulde} one completely ignores the largest
electron hopping amplitude $t_{\perp}$, which would result in the
random exchange between electron spins above $T_c$, instead of
the quasi-one-dimensional behavior.

The results of Seo and Fukuyama \cite{Fukuyama} are closer to
ours; they also concluded that the zigzag structure is more
favorable.  Their arguments, however, are based on nn and nnn
Coulomb terms only, which is definitely insufficient for the
Coulomb interaction (as we mentioned above, the {\em long range}
Coulomb interaction seems to favor the \VFOUR-\VFIVE\ chain
structure). Furthermore, their singlets are formed by
nearest-neighbor pairs between ladders, which, as we argued
above, is rather questionable.


\begin{references}
\bibitem{brick} The resulting effective spin model is rather
interesting in its own right. Topologically it is equivalient to
a ``brick wall'' spin lattice with the ``horizontal'' exchange
constant $J_{hor} \sim t_{||}^2 / \Delta$ (interaction in ladders
between sites A, B, C, $\ldots$ in Fig.~\ref{alter}) and the
``vertical'' exchange, $J_{vert}$,  describing the interaction in
nn interaldder pairs, e.g., C and D in Fig.~\ref{alter}.
\end{references}
\end{document}